\documentclass[twocolumn,showpacs,amsmath,amssymb,floats,floatfix,prb,aps,superscriptaddress,showkeys]{revtex4}

\usepackage{graphicx}% Include figure files
\usepackage{epsfig}
\usepackage{dcolumn}% Align table columns on decimal point
\usepackage{bm}% bold math
\usepackage{times}
\begin{document}
\title{Electrostatic analogy of the Jackiw-Rebbi zero energy state}
\author{Gabriel Gonz\'alez}\email{ggonzalezco@conacyt.mx}
\affiliation{C\'atedras CONACYT, Universidad Aut\'onoma de San Luis Potos\'i, San Luis Potos\'i, 78000 MEXICO}
\affiliation{Coordinaci\'on para la Innovaci\'on y la Aplicaci\'on de la Ciencia y la Tecnolog\'ia, Universidad Aut\'onoma de San Luis Potos\'i,San Luis Potos\'i, 78000 MEXICO}
\author{Javier M\'endez}
\affiliation{Coordinaci\'on para la Innovaci\'on y la Aplicaci\'on de la Ciencia y la Tecnolog\'ia, Universidad Aut\'onoma de San Luis Potos\'i,San Luis Potos\'i, 78000 MEXICO}
\author{Ram\'on Diaz de Le\'on-Zapata}
\affiliation{Instituto Tecnol\'ogico de San Luis Potos\'i, Avenida Tecnol\'ogico s/n, 78376 Soledad de Graciano S\'anchez, SLP, MEXICO}
\affiliation{Coordinaci\'on para la Innovaci\'on y la Aplicaci\'on de la Ciencia y la Tecnolog\'ia, Universidad Aut\'onoma de San Luis Potos\'i,San Luis Potos\'i, 78000 MEXICO}
\author{Francisco Javier Gonz\'alez}
\affiliation{Coordinaci\'on para la Innovaci\'on y la Aplicaci\'on de la Ciencia y la Tecnolog\'ia, Universidad Aut\'onoma de San Luis Potos\'i,San Luis Potos\'i, 78000 MEXICO}
\begin{abstract}
We present an analogy between the one dimensional Poisson equation in inhomogeneous media and the Dirac equation  in one space dimension with a Lorentz scalar potential for zero energy. We illustrate how the zero energy state in the Jackiw-Rebbi model can be implemented in a simple one dimensional electrostatic setting by using an inhomogeneous electric permittivity and an infinite charged sheet. Our approach provides a novel insight into the Jackiw-Rebbi zero energy state and provides a helpful view in teaching this important quantum field theory model using basic electrostatics.
\end{abstract}

\maketitle
\section{Introduction}
The Dirac equation is one of the fundamental equations in theoretical physics that accounts fully for special relativity in the context of quantum mechanics for elementary spin-1/2 particles.\cite{pam} The Dirac equation plays a key role to many exotic physical phenomena such as graphene,\cite{novo} topological insulators\cite{topo} and superconductors.\cite{topos} These systems proved to be ideal testing grounds for theories of the coexistence of quantum and relativistic effects in condensed matter physics.\\
Recently, a significant number of studies has addressed the problem of simulating relativistic quantum mechanics using different physical platforms such as optical structures,\cite{ucf1,gg} metamaterials\cite{wei} and ion traps.\cite{lama} These studies are based on the mathematical analogies found between different physical theories which provides a way to explore at a macroscopic level many quantum phenomena which are currently inaccessible in microscopic quantum systems. Among the wide variety of quantum-classical analogies investigated so far it appears that the most fruitful one is given by the analogy between optics with quantum phenomena due naturally to the duality between matter and optical waves. The study of quantum-optical analogies is based on the formal similarity between the paraxial optical wave equation in dielectric media and the single particle Schrodinger equation.\cite{longhi} Among the wide variety of quantum-optical analogies we can mention the Bloch oscillations and Zener tunneling, dynamic localization, Anderson localization, quantum Zeno effect, Rabi flopping and coherent population trapping. All these progress has led to the area of research of how relativistic quantum systems can be mimic by optical waves. More recently, optical systems governed by the relativistic Dirac equation have been investigated experimentally such as Klein Tunneling, Zitterbewegung and the Jackiw-Rebbi model.\\
The purpose of this article is to demonstrate that electrostatics can provide a laboratory tool where physical phenomena described by the Dirac equation can be explore. In particular, we demonstrate that the Poisson equation in one dimensional inhomogeneous media can be mapped into the zero energy state of the Dirac equation in one dimension with a Lorentz scalar potential. By tailoring the electric permittivity we propose an electrostatic experiment that simulates a historically important relativistic model known as the Jackiw-Rebbi model.\cite{jackreb} Since the derivation of this important model many useful variations of the Jackiw-Rebbi model have been investigated such as the Ramajaran-Bell model,\cite{bell} the massive Jackiw-Rebbi model,\cite{mjr} the coupled fermion-kink model\cite{fk} and the Jackiw-Rebbi model in distinct kinklike backgrounds.\cite{kg}  \\
The article is organized as follows. First we will start with a brief review of the Jackiw-Rebbi model and how one can obtain the zero energy state of the JR model. Then we will show how the Poisson equation can be mapped into
a Dirac-like equation, and illustrate how the zero energy state in the Jackiw-Rebbi model can be implemented in a simple one dimensional electrostatic setting by using an infinite charged sheet separating two different media. The conclusions are summarized in the last section.
\section{Jackiw-Rebbi model in one dimension}
The Jackiw-Rebbi model describes a one dimensional Dirac field coupled to a static background soliton field and is known as one of the earliest theoretical description of a topological insulator where the zero energy mode can be understood as the edge state. In particular, the Jackiw-Rebbi model has been studied by Su, Shrieffer and Heeger in the continuum limit of polyacetylene.\cite{shrieffer} The one dimensional Dirac equation in the presence of an external field $\varphi(x)$ and with $\hbar=c=1$ is given by
\begin{equation}
\hat{H}_D\Psi(x)=\left[\sigma_y \hat{p}+\sigma_x\varphi(x)\right]\Psi(x)={\mathcal E}\Psi(x)
\label{eq1}
\end{equation}
where 
\begin{eqnarray} 
\sigma_y=\left(\begin{array}{cc} 0 & -i\\ i & 0 \end{array}\right),\quad \sigma_x=\left(\begin{array}{cc} 0 & 1\\ 1 & 0 \end{array}\right) \quad \mbox{and} \quad \Psi=\left(\begin{array}{c}\psi_1\\ \psi_2\end{array}\right).
\label{eq2}
\end{eqnarray}
We use the Pauli matrices $\sigma_x$ and $\sigma_y$ in order to have a real two component spinor $\Psi(x)$. From eq. (\ref{eq2}) it follows that the Dirac Hamiltonian possesses a {\it chiral} symmetry defined by the operator $\sigma_z$, which anticommutes with the Dirac Hamiltonian, i.e. $\{\hat{H}_D,\sigma_z\}=0$. The {\it chiral} symmetry implies that eigenstates come in pairs with positive and negative energy $\pm{\mathcal E}$, respectively. It is possible however for an eigenstate to be its own partner for ${\mathcal E}=0$, if this is the case then the state is topologically protected. The resulting zero energy state is protected by the topology of the scalar field, whose existence is guaranteed by the index theorem, which is localised around the soliton.\cite{jackreb}\\
The Jackiw-Rebbi model uses $\varphi(x)=m\tanh{(\lambda x)}$ for the external scalar field, with $m>0$ and $\lambda>0$. For simplicity we will consider a external scalar field given by 
\begin{equation}
\varphi(x)=m\frac{x}{|x|}
\label{eq3}
\end{equation} 
forming a domain wall at $x=0$ where $\varphi(x=0)=0$. The scalar field given by eq. (\ref{eq3}) is a simplification of the Jackiw-Rebbi model first proposed by Rajaraman-Bell.\cite{bell} The precise form of the external scalar potential is not important as long as it asymptotically approaches an opposite sign at $x\rightarrow\pm\infty$. The wave function may change corresponding to a particular form of the external scalar potential, but the existence of the zero energy state is determined solely by the fact that the mass is positive on one side and negative on the other. Therefore, the solution is very robust against the external scalar potential.\\ 
\begin{figure}[t]
\centering
\includegraphics[width=\linewidth]{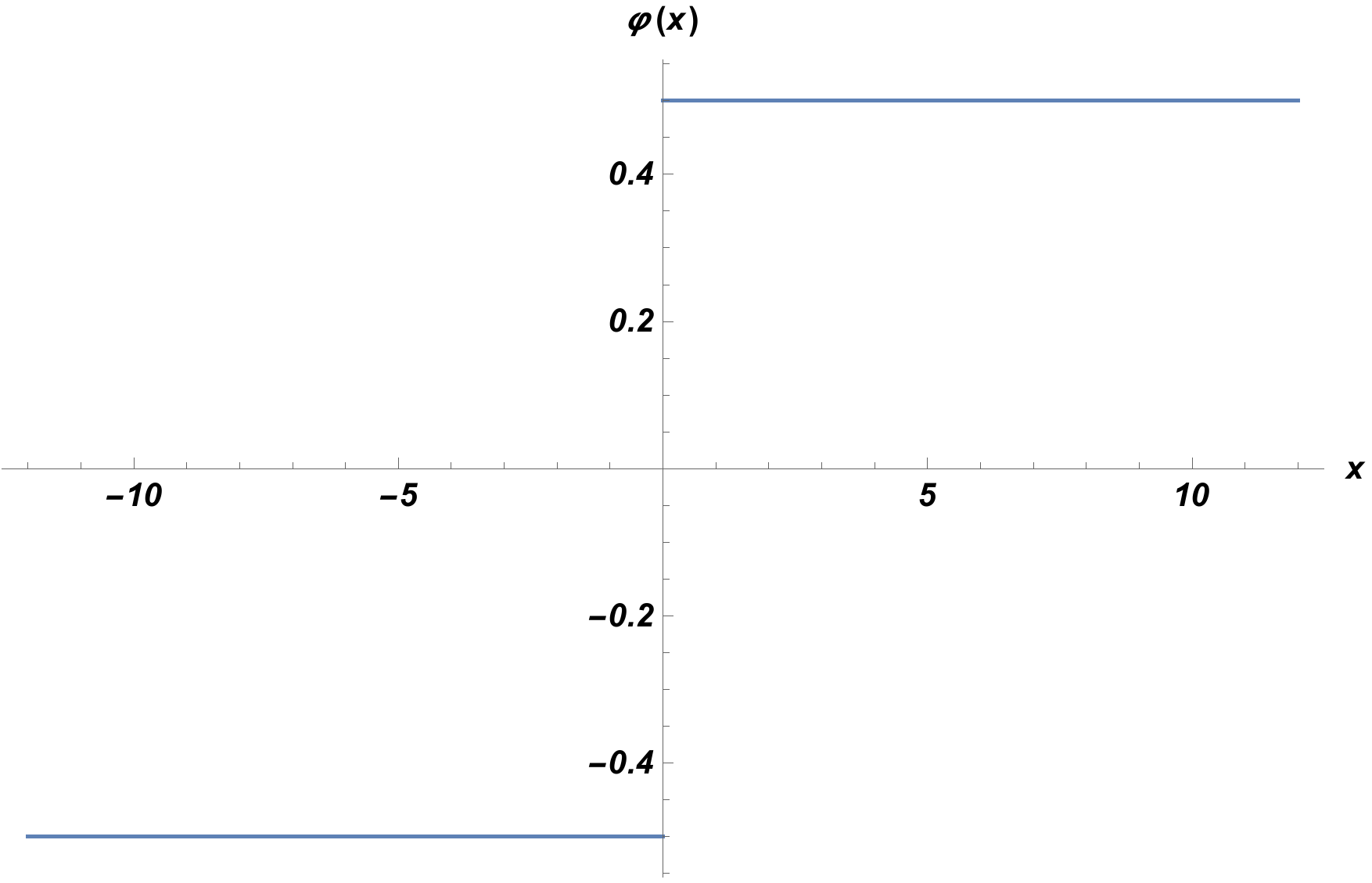} 
\caption{The figure shows the external scalar potential $\varphi(x)$ which changes sign at the interface $x=0$.}
\label{fig1}
\end{figure}
\begin{figure}[h!]
\centering
\includegraphics[width=\linewidth]{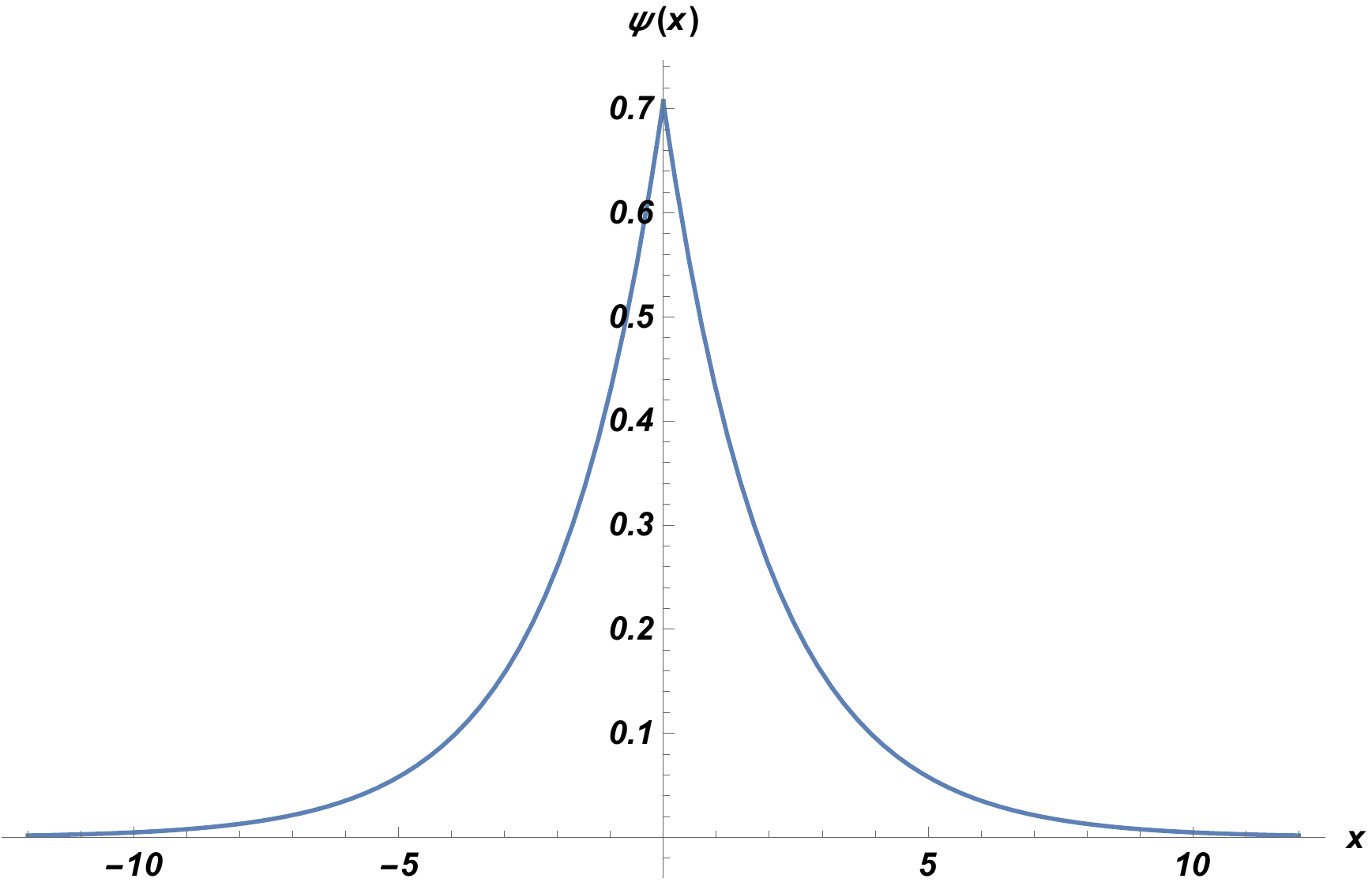} 
\caption{The figure shows the Jackiw-Rebbi zero energy mode given by Eq.(\ref{eq5}) for the external scalar field depicted in Fig.(\ref{fig1}). Note how the zero energy state is localized around the interface $x=0$.}
\label{fig2}
\end{figure}
\\
The solution of the Dirac equation at exactly zero energy for the scalar field given by eq. (\ref{eq3}) is obtained by solving the following equation 
\begin{equation}
\left(\begin{array}{cc}
0 & -\partial_x+\varphi(x)\\ \partial_x+\varphi(x) & 0\end{array} \right)\left(\begin{array}{c}\psi_1(x) \\ \psi_2(x)\end{array}\right)=0 
\label{eq4}
\end{equation} 
which gives
\begin{equation}
\psi_{i}=C_{\mp}\exp\left[\mp m|x|\right], \quad\mbox{for $i=1,2$.}
\label{eq5}
\end{equation}
where $C_{\mp}$ is a normalization constant and the double sign in eq.(\ref{eq5}) is $-(+)$ for $i=1(2)$. Note that $\psi_{1,2}$ cannot be both normalized. If we impose that $\lim_{x\rightarrow\pm\infty}\psi_i(x) \rightarrow 0$ we need to make $C_+=0$ in order to have a properly normalized state. In Fig.(\ref{fig2}) we show the wave function for the zero energy state of the Jackiw-Rebbi model.
\section{Electrostatic analog of the Jackiw-Rebbi model}
In this section we show that the zero energy Jackiw-Rebbi state can be generated at the interface of two dielectric materials separated by a infinite charged sheet. The use of infinite charged sheets for emulating physical systems has been used extensively in the past for a wide range of applications such as a simple parallel plate capacitor\cite{griff} or the study of the one dimensional Coulomb gas.\cite{lenard,gg4,gg5} Since we will be working with a planar charge distribution we will consider only the one dimensional Poisson equation with an inhomogeneous electric permittivity, i.e. $\epsilon(x)$, which is given by 
\begin{equation}
\frac{d}{dx}\left(\epsilon(x)\frac{dV}{dx}\right)=-\rho(x),
\label{eq01}
\end{equation} 
where $V(x)$ is the electrostatic potential and $\rho(x)$ is the volume charge distribution.  Expanding the left hand side of eq. (\ref{eq01}) and multiplying by $1/\epsilon$ we have
\begin{equation}
\frac{d^2V}{dx^2}+\frac{\epsilon^{\prime}}{\epsilon}\frac{dV}{dx}=-\frac{\rho}{\epsilon},
\label{eq02}
\end{equation}
where $\epsilon^{\prime}$ represents the total derivative with respect to the space coordinate $x$. Let us now make the following transformation
\begin{equation}
V(x)=V_0\ln\left(\psi_1(x)/A\right),
\label{eq03}
\end{equation}
where $V_0$ and $A$ are constants to ensure dimensional consistensy, and $\psi_1(x)$ is an arbitrary function. Substituting eq. (\ref{eq03}) into eq. (\ref{eq02}) we have
\begin{equation}
V_0\frac{\psi_1^{\prime\prime}}{\psi_1}-\frac{\epsilon^{\prime}}{\epsilon}E_x-\frac{1}{V_0}E_x^2=-\frac{\rho}{\epsilon}
\label{eq04}
\end{equation}
where we have used the identity $E_x=-dV/dx$. If we use eq. (\ref{eq01}) in the right hand side of eq. (\ref{eq04}) we end up with the following equation
\begin{equation}
V_0\frac{\psi_1^{\prime\prime}}{\psi_1}-\frac{dE_x}{dx}-\frac{1}{V_0}E_x^2=0,
\label{eq04a}
\end{equation}
note that eq. (\ref{eq04a}) does not depend on the electric permittivity anymore. Multiplying eq. (\ref{eq04a}) by $\psi_1/V_0$ and adding and subtracting the term $E_x\psi_1^{\prime}/V_0$ to the left hand side of eq. (\ref{eq04a}) we have
\begin{equation}
\lim_{{\mathcal E}\rightarrow 0}\left\{\left[\psi_1^{\prime}+\frac{E_x}{V_0}\psi_1\right]^{\prime}-\frac{E_x^2}{V_0^2}\psi_1-\frac{E_x}{V_0}\psi_1^{\prime}+{\mathcal E}^2\psi_1\right\}=0
\label{eq05}
\end{equation}
where ${\mathcal E}$ is an auxiliary constant that we will set to zero at the end of our calculations. If we make the following substitution $\psi_1^{\prime}+E_x\psi_1/V_0={\mathcal E}\psi_2$ into eq. (\ref{eq05}),  we end up with the following equation $-\psi_2^{\prime}+E_x\psi_2/V_0={\mathcal E}\psi_1$. These two coupled differential equations can be written in the same mathematical form as the Dirac equation with $c=\hbar=1$, i.e.
\begin{equation}
\hat{H}_D\Psi=\left[\sigma_y \hat{p}+\sigma_x\left(\frac{E_x}{V_0}\right)\right]\Psi={\mathcal E}\Psi.
\label{eq06}
\end{equation}
%where
%\begin{eqnarray} 
%\sigma_y=\left(\begin{array}{cc} 0 & -i\\ i & 0 \end{array}\right),\quad \sigma_x=\left(\begin{array}{cc} 0 & 1\\ 1 & 0 \end{array}\right) \quad \mbox{and} \quad \Psi=\left(\begin{array}{c}\psi_1\\ \psi_2\end{array}\right).
%\label{eq07}
%\end{eqnarray}
Equations (\ref{eq06}) can be reduced to two uncoupled Schr\"odinger equations $\hat{H}_{i}\psi_{i}=0$, for $i=1,2$, given by
\begin{equation}
\hat{H}_i\psi_i=\left(\frac{\partial^2}{\partial x^2}+U_i(x)\right)\psi_i(x)=0
\label{eq08}
\end{equation}
where
\begin{equation}
U_{1,2}(x)=\left[-\left(\frac{E_x}{V_0}\right)^2+{\mathcal E}^2\pm\frac{1}{V_0}\frac{dE_x}{dx}\right].
\label{eq09}
\end{equation}
Clearly, $\hat{H}_{1,2}$ are supersymmetric partner Hamiltonians which can be factorized as 
$\hat{H}_1=\hat{A}^{\dagger}\hat{A}-{\mathcal E}^2$ and $\hat{H}_2=\hat{A}\hat{A}^{\dagger}-{\mathcal E}^2$ where $\hat{A}=(\partial_x+E_x/V_0)$ and $\hat{A}^{\dagger}=(-\partial_x+E_x/V_0)$. The relation between Poisson's equation and Schr\"odinger equation in one dimension has been pointed out before by one of the authors (GG).\cite{gg2,gg3}\\
We can easily construct the zero energy mode by setting ${\mathcal E}=0$ in eq.(\ref{eq06}) and solving for the uncoupled first order differential equations for $\psi_{1,2}$, i.e.
\begin{equation}
\psi_{i}=C_{\mp}\exp\left[\mp\int\left(\frac{E_x}{V_0}\right)dx\right]
\label{eq10}
\end{equation}
where $C_{\mp}$ is a normalization constant and the double sign in eq.(\ref{eq10}) is $-(+)$ for $i=1(2)$. The existence of a zero energy state then depends on the asymptotic behavior of $E_x$.\\
We know from basic electrostatics that the electric field due to an infinite charged sheet with volume charge density $\rho(x)=\sigma\delta(x)$ with $\sigma>0$ separating two dielectric materials with electric permittivity $\epsilon_1$ and $\epsilon_2$ is given by
\begin{equation}
E_x(x)=\left\{\begin{array}{ll}
																									\frac{\sigma}{2\epsilon_1}, \mbox{for $x>0$} \\
																									-\frac{\sigma}{2\epsilon_2}, \mbox{for $x<0$}.
																									\end{array}\right.
\label{eq11}
\end{equation}
Interestingly, the elestrostatic field given by eq.(\ref{eq11}) has the same form as the external scalar field given by eq.(\ref{eq3}) that allows the existence of the zero energy state in the JR model.\\
The electrostatic potential for the electric field given by eq. (\ref{eq11}) is 
\begin{equation}
V(x)=-\int_0^xE_x(x)dx=\left\{\begin{array}{ll}
																									-\frac{\sigma}{2\epsilon_1}x, \mbox{for $x>0$} \\
																									\frac{\sigma}{2\epsilon_2}x, \mbox{for $x<0$}.
																									\end{array}\right.
\label{eq12}
\end{equation}
Using eq. (\ref{eq10}) we see that we need to set $C_+=0$ in order to make the two-component spinor normalizable. Therefore, the normalized wave function for the zero mode is given by
\begin{equation}
\Psi(x)=\sqrt{\frac{\sigma}{V_0(\epsilon_1+\epsilon_2)}}\left(\begin{array}{c} e^{V(x)/V_0} \\ 0\end{array}\right).
\label{eq14}
\end{equation}
\begin{figure}[t]
\centering
\includegraphics[width=\linewidth]{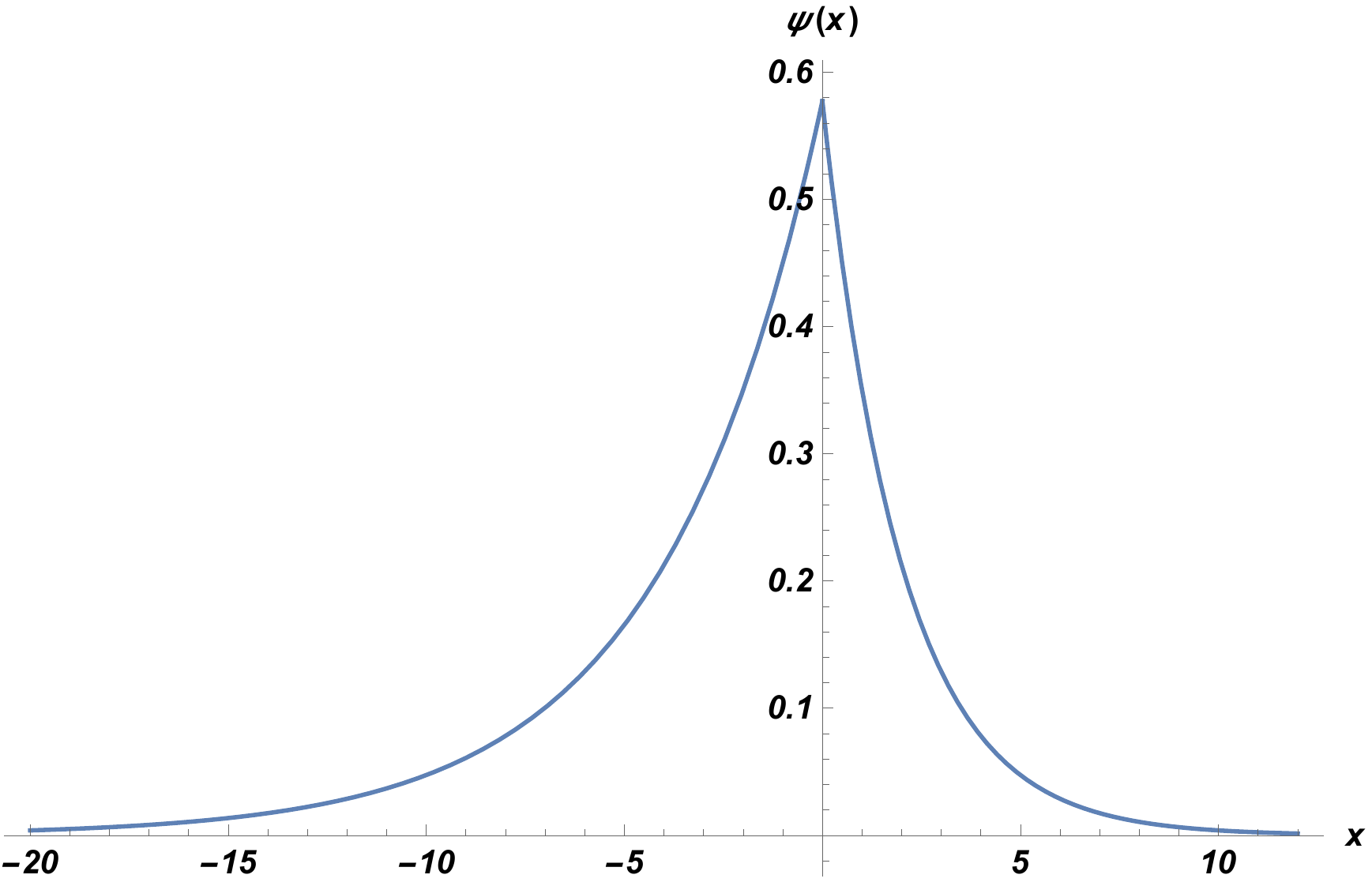} 
\caption{The figure shows the electrostatic Jackiw-Rebbi zero energy mode given by Eq.(\ref{eq14}) for the following values $\sigma=V_0=1$, $\epsilon_1=1$ and $\epsilon_2=2$.}
\label{fig3}
\end{figure}
In Fig.(\ref{fig3}) we show the electrostatic zero energy wave function for the Jackiw-Rebbi model, the wave function dominantly distributes near the interface $x=0$ and decays exponentially away. The solution given in eq. (\ref{eq14}) for $\epsilon_1=\epsilon_2$ is the same as the Jackiw-Rebbi zero energy state.
\section{Conclusions}
In conclusion we have shown that the Poisson equation in one dimensional inhomogeneous media can be used to simulate the Jackiw-Rebbi model in one space dimension for the zero energy state. In particular, we demonstrate how the zero energy state of the Jackiw-Rebbi model can be implemented in an electrostatic set up with an infinite charged sheet that separates two different media. Based on these findings, we have introduced an electrostatic platform for realizing the zero energy state of the Jackiw-Rebbi model which allows one to probe in the laboratory.
\section{Acknowledgments}
This work was supported by the program ``C\'atedras CONACYT". FJG would like to acknowledge support from project 32 of ``Centro Mexicano de Innovaci\'on en Energ\'ia Solar" and by the National Laboratory program from CONACYT through the Terahertz Science and Technology National Lab (LANCYTT).

\end{document}